\documentclass[12pt, reqno]{amsart}
\usepackage{amsmath, amsthm, amscd, amsfonts, amssymb, graphicx, color}
\usepackage[bookmarksnumbered, colorlinks, plainpages]{hyperref}

\begin{document}
\title[THE PRICING OF MULTIPLE-EXPIRY EXOTICS]{THE PRICING OF MULTIPLE-EXPIRY EXOTICS}

\author[HYONG-CHOL O,  MUN-CHOL KIM]{HYONG-CHOL O,  MUN-CHOL KIM\\\\
 Faculty of Mathematics, Kim Il Sung University,  Pyongyang, D. P. R. Korea}

\address{HYONG-CHOL O \newline
Faculty of Mathematics, Kim Il Sung University,  Pyongyang, D. P. R. Korea}
\email{ohyongchol@yahoo.com}

\address{MUN-CHOL KIM  \newline
Faculty of Mathematics, Kim Il Sung University,  Pyongyang, D. P. R. Korea}
\email{ohyongchol@163.com}

\thanks{Last revised Aug 7, 2013, published in {\it Elextronic Journal of Mathematical Analysis and Applications, Vol.1(2) July 2013, pp 247-259}}

\setcounter{page}{1}

\maketitle

\begin{abstract}
In this paper we extend Buchen's method to develop a new technique for pricing of some exotic options with several expiry dates (more than 3 expiry dates) using a concept of higher order binary option. At first we introduce the concept of higher order binary option and then provide the pricing formulae of $n$-th order binaries using PDE method. After that, we apply them to get the pricing of some multiple-expiry exotic options such as Bermudan option, multi time extendable option, multiple shout option and etc. Here, when calculating the price of concrete multiple-expiry exotic options, we do not try to get the formal solution to corresponding initial-boundary problem of the Black-Scholes equation, but explain how to express the expiry payoffs of the exotic options as a combination of the payoffs of some class of higher order binary options. Once the expiry payoffs are expressed as a linear combination of the payoffs of some class of higher order binary options, in order to avoid arbitrage, the exotic option prices are obtained by static replication with respect to this family of higher order binaries.
\end{abstract}

\section{Introduction}
European and American options are referred to as \textit{vanilla options}. Vanilla options have a single future expiry payoff that corresponds to buying or selling the underlying asset for a fixed amount called the \textit{strike price}. European options can only be exercised at the expiry date, whereas American options may be exercised at any time before and at the expiry date. The various needs of risk management in financial markets give rise to many exotic (not ordinary) options with various payoff structures, and thus a lot of exotic options continue to be popular in the over-the counter market. Among them, there exists a class of exotic options whose payoff structure involves \textit{several fixed future dates}, here we write them by $T_{0}<T_1< \cdots
<T_N$. Usually, at the first expiry date $T_0$ , the option holder receives a contract related with dates $T_1, \cdots,T_N$. Options belonging to this class are called \textit{multiple-expiry exotic options}. Bermudan options or several times extendable options are good examples.\\\indent
In this paper we extend Buchen's method \cite{Buc} to develop a new technique for pricing of multiple expiry exotic options in terms of a portfolio of higher order binary options. An \textit{up} (or \textit{down}) binary option is the option that on its expiry date delivers an agreed payoff if the price of the underlying asset is above (or bellow) a fixed exercise price and zero otherwise. Binary options whose agreed payoff is an asset are referred as \textit{first order asset binaries} and binary options whose agreed payoff is cash are referred as \textit{first order bond binaries} \cite{Buc}. In this paper, the binary options whose agreed payoff is a $(n-1)$-th order binary are called $n$-th \textit{order binaries} by induction.\\\indent
The basic idea of expressing the payoffs of complex options in terms of binary options can be seen in previous publications. Rubinstein and Reiner \cite{RR1} considered the relationship of barrier options and binaries. Ingersoll \cite{Ing} extended the idea by expressing complex derivatives in terms of "\textit{event-driven}" binaries. Event $ \varepsilon$-driven binary option pays one unit of underlying asset if and only if the event $\varepsilon$ occurs, otherwise it pays nothing.\\\indent
Buchen \cite{Buc} developed a theoretical framework for pricing dual-expiry options in terms of a portfolio of elementary binary options. He introduced the concepts of first and second order binary options and provided the pricing formula for them using \textit{expectation method}. And then he applied them to pricing some dual expiry exotic options including compound option, chooser option, one time extendable option, one time shout option, American call option with one time dividend and partial barrier options.\\\indent 
The purpose of this paper is to extend the Buchen's procedure to the case of \textit{multiple-expiry options} (more than 3 expiry dates). To do that, here, at first we introduced the concept of \textit{higher order binary options} and then provided the \textit{pricing formula} of $n$-th \textit{order binaries} using \textit{PDE method}. After that, we applied them to pricing of some multiple-expiry exotic options such as Bermudan options, multi time extendable options, multiple shout options and etc.\\\indent 
As Buchen \cite{Buc} mentioned, historically, dual expiry exotics as well as multiple-expiry option prices have been derived individually by various authors. This paper has demonstrated that multiple-expiry options can be priced in a unified framework by expressing each as a portfolio of higher order binaries.\\\indent  
The focus of this paper is on explaining the basic pricing method in terms of static replication with higher order binaries, and so issues such as the financial motivation for trading multiple-expiry exotics, computation and simulation will not be considered here in detail and the explanation about another authors' results on concrete multiple-expiry exotics such as Bermudan option, extendable option and shout option will refer to \cite{Lon, Sch, Tho}. 

\section{Higher Order Binary Options}
Consider an underlying asset (for example, a stock) whose price $x$ satisfies Ito stochastic differential equation. Let $r, q$ and $\sigma$ be respectively risk free rate, dividend rate and volatility. Then to avoid arbitrage, the price of $V(x, t)$ of any derivative on the stock with expiry date $T$ and expiry payoff $f(x)$ must satisfy the following \eqref{1}-\eqref{2}.
\begin{equation}
\frac{\partial V}{\partial t}+\frac{\sigma^{2}}{2} x_2 \frac{\partial^2 V}{\partial t^2}+(r-q)x\frac{\partial V}{\partial x}-rV=0,~0\leq t<T,~0<x<\infty \label{1} 
\end{equation}
\begin{equation}
V(x,~T)=f(x) \label{2}
\end{equation}
This price $V(x,t)$ is called a \textit{\textbf{standard option}} with expiry payoff $f(x)$ \cite{Buc}. \\\indent

{\bf Proposition 1}\cite{Kwo} {\it Assume that there exist non negative constants $M$ and $\alpha$ such that $|f(x)| \leq Mx^{\alpha \ln x},~x>0$. Then the price of standard option, that is, the solution of \eqref{1} and \eqref{2} is provided as follows:}  \newline
\[
V(x, t;T)=e^{-r(T-t)}\int_{0}^{\infty} \frac{1}{\sigma \sqrt{2 \pi (T-t)}} \frac{1}{z}e^{-\frac{[\ln \frac{x}{z}+(r-q- \frac{\sigma^{2}}{2})(T-t)]^2}{2 \sigma^2 (T-t)}}f(z)dz
\]
\begin{equation}
=xe^{-q(T-t)}\int_{0}^{\infty} \frac{1}{\sigma \sqrt{2 \pi (T-t)}} \frac{1}{z^2}e^{-\frac{[\ln \frac{x}{z}+(r-q+ \frac{\sigma^{2}}{2})(T-t)]^2}{2 \sigma^2 (T-t)}}f(z)dz~.  \label{3}
\end{equation}
\\
{\bf Remark.} It is well known that the change of variable $y=\ln x$ transforms \eqref{1} to a parabolic PDE with constant coefficients which can be easily transformed into a heat equation. (For example, see \cite{Jia}.) From the theory of heat equations we can know that the singular integral on the left side of \eqref{3} and its $t$ and $x$ derivatives always exist under the above condition on $f$, which can be easily seen in references such as \cite{RR2} on PDE or equations of mathematical physics, and we can easily check that \eqref{3} satisfies \eqref{1} and \eqref{2}.   
\\

An {\bf \textit{up binary option}} of {\it exercise price} $\xi$ on the {\it standard option} with expiry payoff $f(x)$ is a contract with expiry payoff $f(x)$ if $x>\xi$ and zero otherwise. A {\bf \textit{down binary option}} pays $f(x)$ if $x<\xi$ and zero otherwise. Let $s$ be the sign "$+$" or "$-$". In what follows we use the sign "$+$" and "$-$" as sign indicators for up and down binaries, respectively. Then the expiry payoff functions for up and down binaries can be written in the form 
\[
V_\xi^s(x, T)=f(x)\cdot 1(sx>s \xi).
\]
From this there holds the following {\it parity relation} between the standard option price and the corresponding up and down binaries:
\[
V_\xi^+(x, t)+V_\xi^-(x, t)=V(x, t),~~t<T.
\]
\indent
If $f(x)=x$, the standard option simply pays one unit of the asset at expiry date $T$, and the price of the standard option is $A(x,t;T)=xe^{-q(T-t)}$ for all $t<T$. And the corresponding binaries are the very {\it asset-or-nothing binaries}, here their prices are denoted by $A_\xi^s(x,t;T)$. Then from the above mentioned parity relation, we have 
\[
A_\xi^+(x,t;T)+A_\xi^-(x,t;T)=A(x, t;T)=x\cdot e^{-q(T-t)},~~t<T.
\]
\indent
If $f(x)=1$, the standard option simply pays one unit of cash at expiry date $T$, and the price of the standard option is $B(x,t;T)=e^{-r(T-t)}$ for all $t<T$. And the corresponding binaries are the very {\it cash-or-nothing binaries} (or {\it bond binaries}), here their prices are denoted by $B_\xi^s(x,t;T)$. From the parity relation, we have
\[
B_\xi^+(x,t;T)+B_\xi^-(x,t;T)=B(x, t;T)=e^{-r(T-t)},~~t<T.
\]
\indent
The concept of {\it Q}-option plays a very useful role in the pricing of dual expiry options and, in particular, simplifying the notation of the price formula. (See \cite{Buc}.) Consider a standard contract that pays $f(x)=s(x-K)$ at expiry date $T$; this is a kind of {\it forward contract}, where the holder must buy if $s=+$ (or sell if $s=-$) one unit of underlying asset for $K$ units of cash. Since $f(x)=s[A(x,T;T)-K\cdot B(x,T;T)]$, the price $Q(x,t;T,K)$ of this contract at time $t<T$ is given by
\[
Q(x,t;T,K)=s[A(x,t;T)-K\cdot B(x,t;T)]=s[x\cdot e^{-q(T-t)}-K\cdot e^{-r(T-t)}].
\]
The binary option that pays $f(x)=s(x-K)\cdot 1(sx>s\xi)$ at expiry date $T$ is called a {\bf \textit{first order Q-option}}, their prices denoted by $Q_\xi^s(x,t;T,K)$. If $\xi =K$, then $Q_K^s(x,t;T,K)$ is is the very ordinary European ({\it call} if $s=+$, or {\it put} if $s=-$) option. So $Q_\xi^s(x,t;T,K)$ is called a {\it generalized European option}. These options are {\it more general} in the sense that their exercise price $\xi$ is different from their strike price $K$. The {\it price of first order Q-options} are given as follows:
\begin{equation}
Q_\xi^s(x,t;T,K)=s[A_\xi^s(x,t;T)-K\cdot B_\xi^s(x,t;T)]. \label{4}
\end{equation}
\indent
The asset or nothing binary, bond binary and the first order {\it Q}-option are called the {\bf \textit{first order binaries}}.\cite{Buc} \\

\indent
{\bf Proposition 2}\cite{Buc, Jia}. {\it The prices of asset and bond binary options are provided as follows:} 
\begin{equation}
A_\xi^s(x, t;T)=x\cdot e^{-q(T-t)}N(sd),~~~~~B_\xi^s(x, t;T)=e^{-r(T-t)}N(sd'). \label{5}
\end{equation}
{\it Here $N(x)$ is the accumulated normal distribution function} 
\[
N(x)=(\sqrt{2\pi})^{-1} \int_{-\infty}^{x}e^{-\frac{y^2}{2}}dy,
\]
{\it and $d$, $d'$ are respectively given as follows:}
\[
d=\frac{\ln \frac{x}{\xi}+(r-q+\frac{\sigma^2}{2})(T-t)}{\sigma \sqrt{T-t}},~~~~d'=d-\sigma \sqrt{T-t}.   
\]
\\ \indent    
{\bf Definition 1.}  An {\bf $n$-th \textit{order binary option}} is a binary contract with expiry date $T_0$ on an underlying $(n-1)$-th order binary option. Specifically, the payoff at time $T_0$ has the following form 
\begin{equation}
V(x,T_0)=F_{\xi_1\cdots \xi_{n-1}}^{s_1\cdots s_{n-1}}(x,T_0;T_1,\cdots ,T_{n-1})\cdot 1(s_0x>s_0\xi_0). \label{6}
\end{equation}
Here $F_{\xi_1\cdots \xi_{n-1}}^{s_1\cdots s_{n-1}}(x,T_0;T_1,\cdots ,T_{n-1})$ is the price of the underlying $(n-1)$-th order binary option with expiry time $T_1,\cdots ,T_{n-1}$ at the time $T_0$ and either $F=A$ if the underlying binary is asset binary, $F=B$ for the underlying bond binary or $F=Q$ for the underlying $Q$-option; and $s_0,\cdots, s_{n-1}$ are up-down indicators ($+$ or $-$) at times $T_0,\cdots ,T_{n-1}$  respectively. $\xi_0,\cdots, \xi_{n-1}$ are their corresponding exercise prices. \\\indent 
The prices of these  $n$-th order binary options at time $t<T_0$ are denoted by $F_{\xi_0 \xi_1\cdots \xi_{n-1}}^{s_{0}s_{1}\cdots s_{n-1}}(x,t;T_0,T_1,\cdots ,T_{n-1})$. \\\indent
Then from the definition we have \\

\indent \indent$~Q_{\xi_0 \xi_1\cdots \xi_{n-1}}^{s_{0}s_{1}\cdots s_{n-1}}(x,t;T_0,T_1,\cdots ,T_{n-1},K)$=\\
\indent \indent \indent \indent \indent  \indent$=s_{n-1}[A_{\xi_0 \xi_1\cdots \xi_{n-1}}^{s_{0}s_{1}\cdots s_{n-1}}(x,t;T_0,T_1,\cdots ,T_{n-1})-$ 
\begin{equation}
-K\cdot B_{\xi_0 \xi_1\cdots \xi_{n-1}}^{s_{0}s_{1}\cdots s_{n-1}}(x,t;T_0,T_1,\cdots ,T_{n-1})]. \label{7}
\end{equation}
\indent
Note that the strike price $K$ in the higher order $Q$-binary is effective only at last time $T_{n-1}$. From the {\it parity relation}, we have 
\[
F_{\xi_0 \xi_1\cdots \xi_{n-1}}^{+s_{1}\cdots s_{n-1}}(x,t;T_0,T_1,\cdots ,T_{n-1})+F_{\xi_0 \xi_1\cdots \xi_{n-1}}^{-s_{1}\cdots s_{n-1}}(x,t;T_0,T_1,\cdots ,T_{n-1})=
\]
\[
=F_{\xi_1\cdots \xi_{n-1}}^{s_{1}\cdots s_{n-1}}(x,t;T_1,\cdots ,T_{n-1}).
\]
\indent
If the pricing formulae of the higher order asset binary $A_{\xi_0 \xi_1\cdots \xi_{n-1}}^{s_{0}s_{1}\cdots s_{n-1}}(x,t;$ $T_0,T_1,\cdots,T_{n-1})$ and the higher order bond binary $B_{\xi_0 \xi_1\cdots \xi_{n-1}}^{s_{0}s_{1}\cdots s_{n-1}}(x,t;T_0,$ $T_1,\cdots ,T_{n-1})$ are provided, then the price of the higher order $Q$-binary is easily provided by \eqref{7}.\\\indent
Black-Scholes expressions for higher order binary options involve {\it multidimensional normal distribution function} $N(x_0,x_1,\cdots,x_{n-1};P)$ with {\it zero mean vector} and {\it correlation matrix} $P^{-1}$:
\begin{equation}
N(x_0,x_1,\cdots,x_{n-1};P)=\int_{-\infty}^{x_0}\cdots \int_{-\infty}^{x_{n-1}}\frac{1}{(\sqrt{2\pi})^n}\sqrt{\det{P}}~e^{-\frac{1}{2}y^\mathsf{T}Py}dy. \label{8}
\end{equation}
Here $y^\mathsf{T}=(y_0,y_1,\cdots,y_{n-1}).$ \\\indent
Let define the matrix $A=(a_{ij})_{i,j=0,1,\cdots,n-1}$ related to the expiry dates $T_0,T_1,\cdots,$ $T_{n-1}$ as follows:\\

   \indent \indent$a_{00}=(T_{1}-t)/(T_{1}-T_{0})$,\\

   \indent \indent$a_{n-1,n-1}=(T_{n-1}-t)/(T_{n-1}-T_{n-2})$,\\ 

   \indent \indent$a_{ii}=(T_{i}-t)/(T_{i}-T_{i-1})+(T_{i}-t)/(T_{i+1}-T_{i}),~~1 \leq i \leq n-2$,\\

   \indent \indent$a_{i,i+1}=a_{i+1,i}=-\sqrt{(T_{i}-t)(T_{i+1}-t)}/(T_{i+1}-T_{i}),~~0 \leq i \leq n-2$,\\\\
and another elements are all zero. Then we have
\[
A^{-1}=(r_{ij}); r_{ij}=\sqrt{(T_{i}-t)/(T_{j}-t)},~r_{ji}=r_{ij}, i\leq j, 
\]
\[
\det{A}=\frac{T_{1}-t}{T_{1}-T_{0}}\cdot \frac{T_{2}-t}{T_{2}-T_{1}}\cdot \cdots \cdot \frac{T_{n-1}-t}{T_{n-1}-T_{n-2}}.
\]
Let $s_i=+$ or $-$ (that is, sign indicators and $i = 0, 1, \cdots, n-1$) and define a new matrix by
\[
A_{s_0s_1\cdots s_{n-1}}=(s_{i}\cdot s_{j}\cdot a_{ij})_{i,j=0}^{n-1}.
\]
Then we have
\[
A_{s_0s_1\cdots s_{n-1}}^{-1}=(s_is_jr_{ij}),\quad\det{A_{s_0s_1\cdots s_{n-1}}}=\det{A}.
\]
\\  
\indent {\bf Theorem 1} {\it The price of $n$-th order asset binary and bond binary are provided as follows} :\\

   \indent \indent \indent$A_{\xi_0 \xi_1\cdots \xi_{n-1}}^{s_{0}s_{1}\cdots s_{n-1}}(x,t;T_0,T_1,\cdots ,T_{n-1})=$
\begin{equation}
=x\cdot e^{-q(T_{n-1}-t)}N(s_0d_0,s_1d_1,\cdots,s_{n-1}d_{n-1};A_{s_0s_1\cdots s_{n-1}}), \label{9}
\end{equation}
   \indent \indent $B_{\xi_0 \xi_1\cdots \xi_{n-1}}^{s_{0}s_{1}\cdots s_{n-1}}(x,t;T_0,T_1,\cdots ,T_{n-1})=$
\begin{equation}
=e^{-r(T_{n-1}-t)}N(s_0d'_0,s_1d'_1,\cdots,s_{n-1}d'_{n-1};A_{s_0s_1\cdots s_{n-1}}). \label{10}
\end{equation}
{\it Here}\\\\
\indent \indent \indent \indent \indent \indent {~} $d_i=\frac{\ln \frac{x}{\xi_i}+(r-q+\frac{\sigma^2}{2})(T_i-t)}{\sigma \sqrt{T_i-t}},~~~~$   
\[
d'_i=d_i-\sigma \sqrt{T_i-t},~i=0,1,\cdots,n-1.   
\]
\indent {\bf Proof:}  The cases of $n=1$ and $n=2$ were proved by Buchen in \cite{Buc} using probability theory and can be easily proved using the formula \eqref{3} too. In the case of $n>2$ we will give a sketch of the proof by induction. \\\indent 
We assume that theorem 1 holds in the case of $n-1$. From the definition 1, $A_{\xi_0 \xi_1\cdots \xi_{n-1}}^{s_{0}s_{1}\cdots s_{n-1}}(x,t;T_0,T_1,\cdots ,T_{n-1})$ satisfies \eqref{1} and  
\[
V(x,T_0)=A_{\xi_1\cdots \xi_{n-1}}^{s_{1}\cdots s_{n-1}}(x,T_0;T_1,\cdots ,T_{n-1})\cdot 1(s_0x>s_0 \xi_0).   
\]
Therefore by the formula \eqref{3}, If we let
\[
G(z)= \frac{1}{\sigma \sqrt{2 \pi (T_0-t)}} \frac{1}{z^2}e^{-\frac{[\ln \frac{x}{z}+(r-q+ \frac{\sigma^{2}}{2})(T_0-t)]^2}{2 \sigma^2 (T_0-t)}},
\]
then $A_{\xi_0 \xi_1\cdots \xi_{n-1}}^{s_{0}s_{1}\cdots s_{n-1}}(x,t;T_0,T_1,\cdots ,T_{n-1})$ is provided as follows:\\

   \indent $A_{\xi_0 \xi_1\cdots \xi_{n-1}}^{s_{0}s_{1}\cdots s_{n-1}}(x,t;T_0,T_1,\cdots ,T_{n-1})=$\\

   \indent {~}~ $=x\cdot e^{-q(T_0-t)}\int_{0}^{\infty}G(z)A_{\xi_1\cdots \xi_{n-1}}^{s_{1}\cdots s_{n-1}}(z,T_0;T_1,\cdots ,T_{n-1})\cdot 1(s_0z>s_0 \xi_0)dz$.\\\\
Here $A_{\xi_1\cdots \xi_{n-1}}^{s_{1}\cdots s_{n-1}}(z,T_0;T_1,\cdots ,T_{n-1})$ is the price of the underlying $(n-1)$-th order asset binary option. By induction-assumption, the formula \eqref{9} holds for $A_{\xi_1\cdots \xi_{n-1}}^{s_{1}\cdots s_{n-1}}(z,T_0;T_1,\cdots ,T_{n-1})$. Thus we have \\

   \indent \indent   \indent \indent$A_{\xi_1\cdots \xi_{n-1}}^{s_{1}\cdots s_{n-1}}(z,T_0;T_1,\cdots ,T_{n-1})=$\\
\[
=z\cdot e^{-q(T_{n-1}-T_0)}N(s_1d_1,\cdots,s_{n-1}d_{n-1};A_{s_1\cdots s_{n-1}}).
\]
Substitute this equality into the above singular integral representation and calculate the integral, then we have \eqref{9} for the case of $n>2$. The proof for \eqref{10} is similar. (Proof End)\\\\\indent
The formulae \eqref{7}, \eqref{9} and \eqref{10} give the following {\bf \textit{price of higher order $Q$-option}} :\\ \\
   \indent $Q_{\xi_0 \xi_1\cdots \xi_{n-1}}^{s_{0}s_{1}\cdots s_{n-1}}(x,t;T_0,T_1,\cdots ,T_{n-1},K)=$ \\

   \indent \indent $=s_{n-1}[x\cdot e^{-q(T_{n-1}-t)}N(s_0d_0,s_1d_1,\cdots,s_{n-1}d_{n-1};A_{s_0s_1\cdots s_{n-1}})-$
\begin{equation}
-K\dot e^{-r(T_{n-1}-t)}N(s_0d'_0,s_1d'_1,\cdots,s_{n-1}d'_{n-1};A_{s_0s_1\cdots s_{n-1}})]. \label{11}
\end{equation}
\\
\section{Applications to Multiple-Expiry Exotics}

In this section, we applied the results of previous section to the pricing of some multiple expiry exotics. \\ \\ \indent
{\bf The Static Replication Theorem} {\it If the payoff of an option at expiry time $T_0$ is a linear combination of prices at time $T_0$ of higher order binaries, then its price at all time $t<T_0$ is the combination of the corresponding prices at time $t$ of the higher order binaries}.\\ \\\indent
The static replication theorem can be proved from the uniqueness of solution to the initial value problem of Black-Scholes equation. And we need one lemma that guarantees the monotonousness of the option price on the underlying asset price and provides some estimates about the gradient of the price on the underlying asset price.\\ \\\indent
{\bf Lemma 1} {\it Assume that $f(x)$ is continuous and piecewise differentiable. If $V(x,t)$ is the solution of \eqref{1} and \eqref{2}, then we have}
\[
\min_x \frac{\partial f}{\partial x}\cdot e^{-q(T-t)}\leq \frac{\partial V}{\partial x}(x,t)\leq \max_x \frac{\partial f}{\partial x}\cdot e^{-q(T-t)}.
\]
{\it In particular, if $E=\{z:\frac{\partial f}{\partial x}(z)<\max_x \frac{\partial f}{\partial x}\},~F=\{z:\frac{\partial f}{\partial x}(z)>\min_x \frac{\partial f}{\partial x}\}, |E|>0$ and $|F|>0$, then}
\[
\min_x \frac{\partial f}{\partial x}\cdot e^{-q(T-t)}<\frac{\partial V}{\partial x}(x,t)<\max_x \frac{\partial f}{\partial x}\cdot e^{-q(T-t)},~t<T.
\]
\indent Using the formula \eqref{3} and the assumption of lemma 1, we can easily prove the required results. \\

\subsection{Bermudan Options}
\begin{figure}[h]
   \centering
   \includegraphics[width=0.55\textwidth]{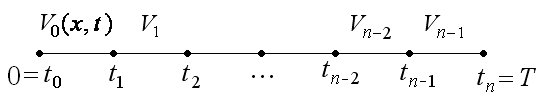}
   \caption{}
  \label{fig1}
\end{figure}

\begin{figure}[h]
\centering
\includegraphics[width=0.53\textwidth]{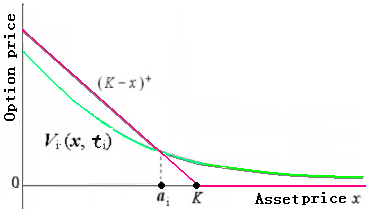}
\caption{}
\label{fig2}
\end{figure}

The {\it Bermudan option} is one type of nonstandard American options and early exercise is restricted to certain dates during the life of the option. For example, let $t_1<t_2<\cdots<t_n=T$ be the dates of early exercise for put option with strike price $K$. Let $t_0=0$. Let denote the option price on the interval $(t_{i-1},t_i]$ by $V_{i-1}(x,t)$, where $i = 1,\cdots,n$. Then \cite{Jia} 
\begin{equation}
V_{i-1}(x,t_i)=\max(V_{i}(x,t_i), (K-x)^{+}),~~i=1,\cdots,n-1.  \label{12}
\end{equation}
(See figure \ref{fig1}.)
\\ \indent
For fixed $i=1,\cdots,n-1$, we consider the following equation:
\begin{equation}
V_{i}(x,t_i)=(K-x)^{+}. \label{13}
\end{equation}
If $V_{i}(x,t_i)$ is monotonously decreasing on $x$, $0<V_i(x,t_i)<K$ and $-1<\frac{\partial V_i(x,t_i)}{\partial x}$, then the equation \eqref{13} has a unique root $a_i$ such that $0<a_i<K$. (See figure \ref{fig2}.)
\\ \indent    
{\bf Lemma 2} {\it Fix $i \in \{1,\cdots,n-1\}$. Assume that $V_i(x,t_i)$ is monotonously decreasing on $x$, $0<V_i(x,t_i)<K$ and $-1<\frac{\partial V_i(x,t_i)}{\partial x}$. Let $a_i$ be the unique root of the equation \eqref{13}. Then for any $t\in [t_{i-1},t_i)$, we have}
\[
V_{i-1}(x,t)=(V_i)_{a_i}^{+}(x,t;t_i)+Q_{a_i}^{-}(x,t;t_i,K)
\]
{\it and $-1\leq \frac{\partial V_{i-1}}{\partial x}\leq 0$. In particular, if $t_{i-1}\leq t<t_i$, then $-1<\frac{\partial V_{i-1}(x,t)}{\partial x}<0$. Here $(V_i)_{a_i}^{+}(x,t;t_i)$ is the solution of Black-Scholes equation \eqref{1} satisfying the condition $V(x,t_i)=V_i(x,t_i)\cdot 1(x>a_i)$}.\\\indent
{\bf Proof.} From the assumption, we can rewrite \eqref{12} as follows. (See figure \ref{fig2}.)
\[
V_{i-1}(x,t_i)=V_i(x,t_i)\cdot 1(x>a_i)+(K-x)\cdot 1(x<a_i).
\]
Thus from the definition of the binary option, we have 
\[
V_{i-1}(x,t)=(V_i)_{a_i}^{+}(x,t;t_i)+Q_{a_i}^{-}(x,t;t_i,K),~t\in (t_{i-1},t_i].
\]
From the assumption we have $0<a_i<K$ and thus using {\it lemma 1}, we have the conclusion on $\frac{\partial V_{i-1}}{\partial x}$. (Proof End)\\\\\indent
Using this lemma, we can easily calculate the price of Bermudan put options. Since the payoff at time $t_n$ is $(K-x)^+$ and we can't early exercise after the time $t_{n-1}$, so the option on the interval $(t_{n-1},t_n]$ becomes an ordinary European put option, and thus its price $V_{n-1}(x,t)$ in the interval $[t_{n-1},t_n)$ is given by 
$$V_{n-1}(x,t)=Q_{K}^{-}(x,t;t_n,K),~t\in [t_{n-1},t_n)$$ 
and $V_{n-1}(x,t_{n-1})$ satisfies $0<V_{n-1}(x,t_{n-1})<K$ and $-1<\frac{\partial V_{n-1}}{\partial x}(x,t_{n-1})$ $<0$. By the {\it static replication theorem} and {\it lemma 2}, our option price at time $t\in [t_{n-2},t_{n-1})$ is given by
\[
V_{n-2}(x,t)=Q_{a_{n-1}K}^{~~~~+~~~~{~~~~-}}(x,t;t_{n-1},t_n,K)+Q_{a_{n-1}}^{~~~-}(x,t;t_{n-1},K),~t\in [t_{n-2},t_{n-1}).
\]
and $-1<\frac{\partial V_{n-2}}{\partial x}<0$. Repeating the similar considerations, we can get the following formulae:\\\\
\indent $V_{i-1}(x,t)=Q_{a_{i}\cdots a_{n-1}K}^{+\cdots {~+}~~~~~~{~~~~~~-}}(x,t;t_i,\cdots,t_n,K)+$ \\\\
\indent \indent $+Q_{a_{i}\cdots a_{n-2}a_{n-1}}^{+\cdots {~+}~~~~~~{~~~~~~~-}}(x,t;t_i,\cdots,t_{n-1},K)+\cdots +Q_{a_{i}a_{i+1}}^{+~~~~{~~~~-}}(x,t;t_i,t_{i+1},K)+$\\
\begin{equation}
+Q_{a_i}^{-}(x,t;t_i,K), ~t\in [t_{i-1},t_{i}),i\in \{1,\cdots,n\}.
\end{equation}\\
In particular, $V_0$ gives the {\bf \textit{price of Bermudan put option}} at time $t=0$.\\\\
\indent $V_{0}(x,t)=Q_{a_{1}\cdots a_{n-1}K}^{+\cdots {~+}~~~~~~{~~~~~~-}}(x,t;t_1,\cdots,t_n,K)+$\\\\
\indent \indent \indent $+Q_{a_{1}\cdots a_{n-2}a_{n-1}}^{+\cdots {~+}~~~~~~{~~~~~~~-}}(x,t;t_1,\cdots,t_{n-1},K)+\cdots +Q_{a_{1}a_{2}}^{+~~{~~-}}(x,t;t_1,t_2,K)+$\\\\
\indent \indent \indent $+Q_{a_1}^{-}(x,t;t_1,K), {~}~{~}~{~}~t\in [0,t_1).$\\
\\
\subsection{Multiple Extendable Options}
\begin{figure}[h]
   \centering
   \includegraphics[width=0.6\textwidth]{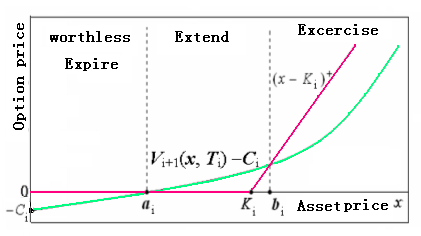}
   \caption{}
   \label{fig3}
\end{figure}

Extendable options were first studied and analyzed in detail by Longstaff \cite{Lon}. Here we consider the {\it holder extendable options} in his sense. The {\it holder} {\bf \textit{$n$-times extendable call}} (or {\bf \textit{put}}) option has the right at some certain dates $T_i$ to exercise a standard European call (or put) option of strike price $K_i$ or to extend the expiry date to time $T_{i+1}(>T_i)$ and change the strike price from $K_i$ to $K_{i+1}(i = 0,1,\cdots,n-1)$ for a premium $C_i$. Here we consider $n$-times extendable call option.\\\indent
Let denote the option price on the interval $[T_{i-1},T_i)$ by $V_i(x,t)$, where $i =0,1,\cdots,n$ and $T_{-1}=0$. Then from the definition of extendable option, we have
\begin{equation}
V_{i}(x,T_i)=\max(V_{i+1}(x,T_i)-C_i, (x-K_i)^{+}),~~i=0,\cdots,n-1.  \label{15}
\end{equation}
\indent
First, let consider the price $V_n(x,t)$ in the last interval $(T_{n-1},T_n]$. In this interval we can no longer extend our option contract and thus our option is just an ordinary European call with the expiry date $T_n$ and strike price $K_n$. Therefore 
\[
V_n(x,t)=Q_{K_n}^{+}(x,t;T_n,K), t\in [T_{n-1},T_n).
\]
And it satisfies $0<\frac{\partial V_{n}(x,T_{n-1})}{\partial x}<1$ by {\it lemma 1}.\\\indent
Now consider the price $V_{n-1}(x,t)$ in the interval $(T_{n-2},T_{n-1}]$. Then by \eqref{15} we have
\[
V_{n-1}(x,T_{n-1})=\max(Q_{K_n}^{+}(x,T_{n-1};T_n,K_n)-C_{n-1}, (x-K_{n-1})^{+}).
\]
From the property of $Q_{K_n}^{+}(x,T_{n-1};T_n,K_n)$ , the following two equations on $x$ have unique roots $a_{n-1}$ and $b_{n-1}$, respectively. \\\\
\indent \indent \indent \indent \indent ${~}~Q_{K_n}^{+}(x,T_{n-1};T_n,K_n)-C_{n-1}=0,$ 
\begin{equation}
Q_{K_n}^{+}(x,T_{n-1};T_n,K_n)-C_{n-1}=(x-K_{n-1})^{+}.
\end{equation}
And it is quite natural that we assume that $a_{n-1}<K_{n-1}<b_{n-1}$. (Otherwise the extension of  $T_{n-1}$ to $T_n$ would not occur as Buchen \cite{Buc} mentioned. See figure \ref{fig3}. Similarly, in put we assume that $a_{n-1}>K_{n-1}>b_{n-1}$.)  
\\
Then we can rewrite $V_{n-1}(x,T_{n-1})$ as follows:\\\\
 \indent $V_{n-1}(x,T_{n-1})=[Q_{K_n}^{+}(x,T_{n-1};T_n,K_n)-C_{n-1}]\cdot 1(a_{n-1}<x<b_{n-1})$\\\\
 \indent \indent \indent \indent  \indent $+(x-K_{n-1})\cdot 1(x>b_{n-1})=$\\\\
 \indent \indent \indent \indent $=Q_{K_n}^{+}(x,T_{n-1};T_n,K_n)\cdot 1(x>a_{n-1})-$\\\\
\indent \indent \indent \indent \indent$-Q_{K_n}^{+}(x,T_{n-1};T_n,K_n)\cdot 1(x>b_{n-1})-$\\\\
 \indent \indent \indent  \indent \indent $-C_{n-1}\cdot 1(x>a_{n-1})+(x-K_{n-1}+C_{n-1})\cdot 1(x>b_{n-1})$\\\\
Thus using the {\it Static Replication Theorem}, for $t\in [T_{n-2},T_{n-1})$ we have\\\\ 
 \indent $V_{n-1}(x,t)=[Q_{a_{n-1}K_n}^{~~{+}~~~{~~~~+}}(x,t;T_{n-1},T_n,K_n)-Q_{b_{n-1}K_n}^{~~+~~{~~~~+}}(x,t;T_{n-1},T_n,K_n)$\\\\
 \indent \indent \indent \indent \indent $-C_{n-1}\cdot B_{a_{n-1}}^{~~{+}}(x,t;T_{n-1})+Q_{b_{n-1}}^{~~{+}}(x,t;T_{n-1},K_{n-1}-C_{n-1})$.\\\\
And $V_{n-1}(x,T_{n-1})$ is a monotone increasing function of $x$ with a positive inclination less than 1 in $x\in (a_{n-1},b_{n-1})$. (See figure \ref{fig3}.) By {\it lemma 1}, we have 
$$0<\frac{\partial V_{n-1}(x,t)}{\partial x}<1,~t\in [T_{n-2},T_{n-1}).$$\\ 
    By similar consideration and induction, we can prove that the formula of price $V_i(x,t)$ of our option in the time interval $[T_{i-1},T_i)$ is provided by 
\[
\sum _{(j_i\cdots j_{n-1})\in \{a_i,b_i\}\times \cdots \times \{a_{n-1},b_{n-1}\}}s(j_i)\cdots s(j_{n-1})\cdot Q_{j_i\cdots j_{n-1}K_n}^{{+}\cdots {~~+}~~{~~+}}(x,t;T_i,\cdots,T_n,K_n)+
\]
\[
\sum _{k=i}^{n-1} \sum _{(j_i\cdots j_{k-1})\in \{a_i,b_i\}\times \cdots \times \{a_{k-1},b_{k-1}\}}s(j_i)\cdots s(j_{k-1})\cdot Q_{j_i\cdots j_{k-1}b_k}^{{+}\cdots {~~+}~~{~~+}}(x,t;T_i,\cdots,T_k,K_k-C_k)
\]
\begin{equation}
+\sum _{k=i}^{n-1}C_k \sum _{(j_i\cdots j_{k-1})\in \{a_i,b_i\}\times \cdots \times \{a_{k-1},b_{k-1}\}}s(j_i)\cdots s(j_{k-1})\cdot B_{j_i\cdots j_{k-1}a_k}^{{+}\cdots {~~+}~~{~~+}}(x,t;T_i,\cdots,T_k).
\end{equation}
Here $s(j)$ is the $a$ or $b$ indicator, that is,
\begin{equation*}
s(j_m)=
\begin{cases}
     1, & j_m=a_m,\\
   -1, &  j_m=b_m.
\end{cases}
\end{equation*}
In particular $V_0$ gives the {\bf \textit{price of n-times extendable call}} option in the time interval $[0,T_0)$.
\\
\subsection{Multiple Shout Options}
Shout options (Thomas \cite{Tho}) are exotic options that allow the holder to lock in a payoff at times prior to the final expiry date. If the shout times would be selected randomly according to the holder's mind, the pricing of shout option will be quite challenging, but if the shout times are pre-determined, then it would be simpler (\cite{Buc}). In a {\it fixed time multiple shout call} option with expiry date $T$ and strike price $K$, their payoff can be locked in at some {\it predetermined} times $T_0<\cdots<T_{n-1}(<T)$. So at the final expiry date $T=T_n$, its payoff is given as follows:
\begin{equation}
V(x,T)=\max (x_0-K,\cdots,x_{n-1}-K,x-K,0), \label{18}
\end{equation}
where $x_i~(i=0,\cdots,n-1)$ are the underlying asset price $x(T_i)$ at the shout time $T_i$. \\
For simplicity, here we consider the case $n=2$. Then the expiry payoff \eqref{18} is given by 
\begin{equation}
V(x,T_2)=\max (x_0-K,x_1-K,x-K,0), \label{19}
\end{equation}
\indent
Note that in the last interval $(T_1,T_2]$ both of the underlying asset prices $x_0=x(T_0)$ and $x_1=x(T_1)$ are known constants and  in the interval $(T_0,T_1]$ the underlying asset price $x_0=x(T_0)$ is a known constant.\\\indent
{\bf The case} of $x_0<K$. We can rewrite \eqref{19} as 
\[
V(x,T_2)=\max (x_1-K,x-K,0)
\]
and this is the terminal payoff of one-shout option in the time interval $(T_0,T_2]$ and thus by the method of \cite{Buc}, for $T_0\leq t<T_1$ we have
\[
V(x,t)=Q_{KK}^{{-}+}(x,t;T_1,T_2,K)+e^{-r(T_2-T_1)}Q_K^{+}(x,t;T_1,K)+
\]
\[
+g(T_1,T_2)A_K^{+}(x,t;T_1),~{~}~{~}T_0\leq t<T_1.
\]
Here
\[
g(T_1,T_2)=e^{-q(T_2-T_1)}N(d^+(T_1,T_2)) - e^{-r(T_2-T_1)}N(d^{-}(T_1,T_2)),
\]
\begin{equation}
d^\pm(T_1,T_2)=[(r-q)/\sigma \pm \sigma /2]\sqrt{T_2-T_1}.  \label{20}
\end{equation}
In particular $x_0=x$ at the time $t=T_0$, we have 
\[
V(x,T_0)=Q_{KK}^{{-}+}(x,T_0;T_1,T_2,K)+e^{-r(T_2-T_1)}Q_K^{+}(x,T_0;T_1,K)+
\]
\begin{equation}
+g(T_1,T_2)A_K^{+}(x,T_0;T_1),~~~~~~~x<K. \label{21}
\end{equation}
\indent  
{\bf The case} of $x_0>K$. If $x_0\leq x_1$, then we can rewrite \eqref{19} as follows:
\[
V(x,T_2)=\max (x_1-K,x-K,0)=(x_1-K)+(x-x_1)^{+}.
\]
Thus in the time interval $[T_1,T_2)$ we have
\[
V(x,t)=(x_1-K)\cdot e^{-r(T_2-t)}+Q_{x_1}^{+}(x,t;T_2,x_1).
\]
In particular, $x_1=x$ at the time $t=T_1$ and 
\[
V(x,T_1)=(x-K)\cdot e^{-r(T_2-T_1)}+Q_{x}^{+}(x,T_1;T_2,x),~~x\geq x_0.
\]
Here by \eqref{4} and \eqref{5} we have
\[
Q_{x}^{+}(x,T_1;T_2,x)=x\cdot g(T_1,T_2),
\] 
where $g(T_1,T_2)$ is as in \eqref{20}. Thus
\begin{equation}
V(x,T_1)=x\cdot [e^{-r(T_2-T_1)}+g(T_1,T_2)] - K\cdot e^{-r(T_2-T_1)},~~x\geq x_0. \label{22}
\end{equation}
If $x_0>x_1$, then we can rewrite \eqref{19} as follows:
\[
V(x,T_2)=\max (x_0-K,x-K,0)=(x_0-K)+(x-x_0)^{+}.
\]
Thus in the time interval $[T_1,T_2)$ we have
\[
V(x,t)=(x_0-K)\cdot e^{-r(T_2-t)}+Q_{x_0}^{+}(x,t;T_2,x_0).
\]
In particular, at the time $t=T_1$ 
\begin{equation}
V(x,T_1)=(x_0-K)\cdot e^{-r(T_2-T_1)}+Q_{x_0}^{+}(x,T_1;T_2,x_0),~~x<x_0. \label{23}
\end{equation}
Combining \eqref{22} with \eqref{23} to get 
\[
V(x,T_1)=[e^{-r(T_2-T_1)}+g(T_1,T_2)]\cdot x\cdot 1(x>x_0)  - e^{-r(T_2-T_1)}\cdot K\cdot 1(x>x_0) 
\]
\[
+(x_0-K)\cdot e^{-r(T_2-T_1)}\cdot 1(x<x_0)+Q_{x_0}^{+}(x,T_1;T_2,x_0)\cdot 1(x<x_0).
\]
Note that in the interval $(T_0,T_1]$ the underlying asset price $x_0=x(T_0)$ is a known constant. Then using the {\it Static Replication Theorem}, for $t\in [T_0,T_1)$, we have 
\[
V(x,t)=[e^{-r(T_2-T_1)}+g(T_1,T_2)]\cdot A_{x_0}^{+}(x,t;T_1)  - K\cdot e^{-r(T_2-T_1)}\cdot B_{x_0}^{+}(x,t;T_1)+ 
\]
\indent \indent \indent \indent \indent $+(x_0-K)\cdot e^{-r(T_2-T_1)}\cdot B_{x_0}^{-}(x,t;T_1)+Q_{x_0x_0}^{{-}+}(x,t;T_1,T_2,x_0)$\\\\
\indent \indent \indent $=[e^{-r(T_2-T_1)}+g(T_1,T_2)]\cdot A_{x_0}^{+}(x,t;T_1)  - K\cdot e^{-r(T_2-t)}+$ \\

\indent \indent \indent \indent $+x_0\cdot e^{-r(T_2-T_1)}\cdot B_{x_0}^{-}(x,t;T_1)+Q_{x_0x_0}^{{-}+}(x,t;T_1,T_2,x_0)$.\\\\
In particular, $x_0=x$ at the time $t=T_0$ and 
\[
V(x,T_0)=[e^{-r(T_2-T_1)}+g(T_1,T_2)]\cdot A_{x}^{+}(x,T_0;T_1)  - K\cdot e^{-r(T_2-T_0)} 
\]
\[
+x\cdot e^{-r(T_2-T_1)}\cdot B_{x}^{-}(x,T_0;T_1)+Q_{x~x}^{-+}(x,T_0;T_1,T_2,x).
\]
Here by \eqref{4} and \eqref{5}, we have
\[
A_{x}^{+}(x,T_0;T_1)= xe^{-q(T_1-T_0)}N(d^{+}(T_0,T_1)),
\]
\[
B_{x_0}^{-}(x,T_0;T_1)=e^{-r(T_1-T_0)}N(d^{-}(T_0,T_1)),
\]
\[
Q_{x~x}^{-+}(x,T_0;T_1,T_2,x)=g_1(T_0,T_1,T_2)\cdot x,
\] 
where $d^{\pm}(T_0,T_1)$ is as in \eqref{20} and 
\[
g_1(T_0,T_1,T_2)=e^{-q(T_2-T_0)}\cdot N_2(-d^+(T_0,T_1),d^+(T_0,T_2);A_{-+})-
\]
\begin{equation}
-e^{-r(T_2-T_0)}\cdot N_2(-d^{-}(T_0,T_1),d^{-}(T_0,T_2);A_{-+}). \label{24}
\end{equation}
Thus we have
\begin{equation}
V(x,T_0)=G(T_0,T_1,T_2)\cdot x - K\cdot e^{-r(T_2-T_0)},~~~x>K. \label{25}
\end{equation}
Here 
\[
G(T_0,T_1,T_2)=[e^{-r(T_2-T_1)}+g(T_1,T_2)]\cdot e^{-q(T_1-T_0)}\cdot N(d^+(T_0,T_1))+
\]
\begin{equation}
+e^{-r(T_2-T_0)}\cdot N(-d^{-}(T_0,T_1))+g_1(T_0,T_1,T_2). \label{26}
\end{equation}
Putting the two expressions \eqref{21} and \eqref{25} of $V(x,T_0)$ together, we have\\\\
\indent \indent \indent $V(x,T_0)=Q_{KK}^{{-}+}(x,T_0;T_1,T_2,K)1(x<K)+$\\\\
\indent \indent \indent \indent \indent $+e^{-r(T_2-T_1)}Q_K^{+}(x,T_0;T_1,K)1(x<K)+$\\\\
\indent \indent \indent \indent \indent  $+g(T_1,T_2)A_K^{+}(x,T_0;T_1)1(x<K)+G(T_0,T_1,T_2)x1(x>K)$\\\\
\indent \indent \indent \indent \indent  $- Ke^{-r(T_2-T_0)}1(x>K)$.\\\\ 
Thus the {\bf \textit{price of fixed time twice shout call option}} at $t<T_0$ is given by\\\\
\indent \indent $V(x,t)=Q_{KKK}^{{-}{-}+}(x,t;T_0,T_1,T_2,K)+e^{-r(T_2-T_1)}Q_{KK}^{{-}+}(x,t;T_0,T_1,K)$\\\\
\indent \indent \indent \indent \indent \indent \indent  $+g(T_1,T_2)A_{KK}^{{-}+}(x,t;T_0,T_1)+G(T_0,T_1,T_2)A_K^{+}(x,t;T_0)$\\
\begin{equation}
- Ke^{-r(T_2-T_0)}B_K^{+}(x,t;T_0),~~~~~~t<T_0. \label{27} 
\end{equation}
\\
\section{Conclusions}
In this paper we introduced the concept of {\it higher order binary options} and then provide the {\it pricing formulae} of $n$-th order binaries using solving method of PDE. Then we applied them to pricing of some {\it multiple-expiry exotic options} such as {\it Bermudan options, multi time extendable options, fixed time twice shout options} and etc. Here when calculating the price of concrete multiple-expiry exotic options, the focus of discussion was on explaining how to express the expiry payoffs of the exotic option as a combination of the payoffs of some class of higher order binary options. Here we assumed that risk free rate, dividend rate and volatility are constant but we could easily extend to the case with time dependent coefficients.\\



\begin{thebibliography}{99}
\bibitem{BY} Broadie, M. and Yamamoto Y., \textit{Application of the Fast Gauss Transform to option pricing}, Management science, Vol. \textbf{49}, No. \textbf{8}, 1071-1088, 2003

\bibitem{Buc} Buchen, P., \textit{The Pricing of dual-expiry exotics}, Quantitative Finance, \textbf{4}, 101-108, 2004

\bibitem{Ing} Ingersoll, J. E., \textit{Digital contract: simple tools for pricing complex derivatives}, J. Business, \textbf{73}, 67-88, 2000

\bibitem{Jia} Jiang, Li-shang, Mathematical Modeling and Methods of Option Pricing, World Scientific, Singapore, 2005

\bibitem{Kwo} Kwok, Y.K., Mathematical models of Financial Derivatives, Springer-verlag, Berlin, 55-82, 1999
 
\bibitem{Lon} Longstaff, F., \textit{Pricing options with extendable maturities: analysis and applications}, J. Finance, \textbf{45}, 935-957, 2000.

\bibitem{RR1} Rubinstein, M. and Reiner, E., \textit{Unscrambling the binary code}, Risk Mag. \textbf{4}, 75-83, 1991

\bibitem{RR2} Rubinstein, I. and L. Rubinstein, Partial Diﬀerential Equations in Classical Mathematical Physics, Cambridge Univ. Press, 296-324, 1998

\bibitem{Sch} Schweizer, M., \textit{On Bermudan Options, Advances in Finance and Stochastics}, in Essays in Honor of Dieter Sondermann, Springer, 257-269, 2002

\bibitem{Tho} Thomas, B., \textit{Something to shout about}, Risk Mag. \textbf{6}, 56-58, 1994

\end{thebibliography}
\end{document}